# Electrotactile vision substitution for 3D trajectory following

A. Chekhchoukh, M. Goumidi, N. Vuillerme, Y. Payan, N. Glade

*Abstract*— Navigation for blind persons represents a challenge for researchers in vision substitution. In this field, one of the used techniques to navigate is guidance. In this study, we develop a new approach for 3D trajectory following in which the requested task is to track a light path using computer input devices (keyboard and mouse) or a rigid body handled in front of a stereoscopic camera. The light path is visualized either on direct vision or by way of a electro-stimulation device, the Tongue Display Unit, a 12x12 matrix of electrodes. We improve our method by a series of experiments in which the effect of the modality of perception and that of the input device. Preliminary results indicated a close correlation between the stimulated and recorded trajectories.

## I. INTRODUCTION

Humans can perceive the real world by way of many senses, vision being one of the most important. Vision is composed of many parts, the eyes being only the peripheral input organ. Despite of eyes deficiencies the brain of a blind person does not lose its capacity to "see". *Visual substitution* aims to replace the absent, defective or occupied visual organ (*e.g.* the eyes) by an artificial sensor (*e.g.* a digital camera). A treatment unit converts the visual information (*i.e.* the image captured by a camera) into a physical stimulation that is exploited by another healthy sensory modality (*e.g.* the tactilo-kinesthetic modality, with the perception of an electro-tactile or vibro-tactile "image") and then understood by the Central Nervous System.

By the end of the 60's the first tactile visual substitution system (TVSS) was designed by Paul Bach-y-Rita. It was composed of a large vibro-tactile matrix (20x20) placed on the chest or on the abdomen of the subject, and coupled with a video camera that captured the dynamical scene. Images were sent to the vibro-tactile matrix in adapted intensity and in a reduced resolution [1]. For many reasons (electrical consumption of the device, miniaturization, and sensory characteristics of the human body) Bach-y-Rita and colleagues converged towards electro-tactile feedback devices using the tongue to convey information. An electro-stimulation device, the Tongue Display Unit or TDU, was then designed [2]. This device consists in a matrix of electrodes put in contact with the superior surface of the tongue and connected to an external central unit. TDU and similar devices are principally used in researches about psycho-motricity, but they are often presented (and sold) as promising devices to give back vision to blind people.

A. Chekhchoukh (abdessalem.chekhchoukh@imag.fr), N. Vuillerme (nicolas.vuillerme@agim.fr) and, N. Glade (nicolas.glade@agim.fr) are with the Univ. Joseph Fourier, CNRS FRE 3405, AGIM Laboratory, Faculty of Medicine of Grenoble, 38700 La Tronche.

M. Goumidi (Malik.goumidi@gmail.com) and Y. Payan (yohan.payan@imag.fr) are with Univ. Joseph Fourier, CNRS UMR 5525, TIMC-IMAG, Grenoble, F-38041, France.

Despite the powerful use of TDU to study substitution paradigms, we do not think such a device can be daily used by blind people. First, because, except for punctual uses, we doubt that those people will accept such a cumbersome and noticeable object that does not allow to talk while used. Second, because the TDU resolution is too low and the sensitivity and the power of discrimination of substitution organs (skin, tongue) are not adapted to allow a good vision substitution, in particular in the case of real complex visual scenes (e.g. scenes of the current life in a street or in a building where many static and moving objects are present). Prosthetic vision seems promising since it is more discrete than the TDU although invasive (since the stimulation matrix is applied directly on the retina). However, currently limited to resolutions of about 10x10 electrodes, it may suffer from the same limitations as the TDU : real complex scenes of current life (with large differences/contrasts in pixels intensities) will appear most of the time as homogenous blots of pixels. Therefore, before technological progress will provide prostheses with higher resolutions, vision substitution with a TDU still constitutes a good way to test different processes to bypass the limitations inherent to these systems. These processes might be applied efficiently afterward in prosthetic vision systems.

There are several other manners to provide information on the environment to blind people via electro-stimulating devices. The simplest is a guidance mode where directional information is given to the subject by way of orders ("go to the left", "go to the right" …) or information about the deviation from a pre-defined trajectory ("you are too much to the left" …), the information being encoded in the form of an activation of several electrodes on the 4 sides of the electro-stimulation matrix. Recently, our team showed that one can efficiently use the TDU in diverse sensori-motor tasks such as (i) guiding a surgical gesture [3, 4], (ii) force production task [5], (iii) joint position sense task [6], (iv) regulating postural sway in bipedal posture [7], or (v) controlling the interface pressure distribution in sitting position [8]. Coupled to geo-location or automatic scene identification, this could be used to guide someone along a recorded trajectory in a certain environment. However, this would not allow the subject to move autonomously in this environment. Another way that gathers autonomous navigation and a good and understandable visual rendering of the environment is to give back to the subject a simplified representation of that environment.

Tatur et al [9] use the stereoscopic information of two video cameras to display image and light information at different distances from the user on the matrix of a mounted head display simulating prosthetic vision. The image switches periodically from short distances to distances far from the user ("sonar" mode) in such a way the user is able to have a representation of the environment along the depth axis. Although it simplifies the information of the images displayed on a stimulation matrix as compared to a classical display of a video stream, this stereoscopic information is probably not the most efficient way to make the visual environment understandable. Indeed, the user perceives successively different images that correspond to different depths, which requires a complex interpretation. Moreover, even if the information is selected (depth and light information only), a visual scene can remain complex and difficult to understand.

We propose a slightly different approach based on the graphical rendering of a model of the environment. Most of the environments, indoor and outdoor, are composed of buildings with window, halls and doors, streets and objects (outside bins, street and traffic lights…). These components can be modeled by a wire-frame scene where the edges would appear in a 3D reconstruction [10, 11]. All component edges are not essential for a blind person to navigate in a secure way. Indeed, only the most relevant paths that can be provided to the person. We propose to display these paths in 3 dimensions in subjective view in the form of a series of light points. Additional elements displayed as static or dynamic symbols will provide information of direction, danger … Coupled to geo-location and postural detection systems, such a process, eventually adapted in prosthetic vision, may constitute an efficient autonomous navigation device. In this article, we present a set-up to improve this concept, in which a user is asked to follow a 3 dimensional pathway to perform a gesture with a tool localized in 3D, the view of the user corresponding to that of the tool.

## II. MATERIAL & METHODS :

### A. Components of the Experimental platform

The experimental platform is composed of (1) a display unit being a 12x12 TDU or (2) a 21" screen (Visual Display Unit or VDU with 128 gray scales) depending on the vision mode used in the experiment. As concerns the recording of the trajectory decided by the subject, two devices are used: (1) a stereoscopic infra-red tracking camera (Polaris, Northern Digital Inc.) coupled with two rigid bodies (one fixed, serving as a reference for position and orientation, and the other one moving along the trajectory thus providing 3D coordinated of the tool) or (2) a keyboard and mouse that can be used instead of rigid bodies to move along the paths.

The platform provides two modes of vision: a visual mode (VDU) and a electro-tactile one (TDU). The TDU is composed of a 3cm matrix of 12x12 electrodes of 1.5mm diameter and spaced each other by 2.3mm. The electro-tactile stimuli are delivered to the dorsum part of the tongue. Voltages are comprised in a range of 1 to 10 volts. A calibration matrix is applied on the image signal to get a uniform and comfortable perception on the overall surface of the tongue [12, 13]. The VDU displays on the screen a graphical larger version of what is sent to the TDU, a visual matrix of 8cm containing 12x12 round points of 2mm diameter.

### B. Paths and tracking

The pathways to follow (Fig. 1) are displayed in 12x12 on the TDU or the VDU in the form of a path of light points that join a start to a target. The environment that contains this light path is viewed in subjective view from the current position (which can be either the tip of the tracked tool with a view along its long axis or a virtual camera whose position and orientation are controlled by the keyboard and the mouse). In the physical world, the light points describe a path from the top to the bottom of a virtual cube of 12cm side. To conserve all the degrees of freedom of the needle (here, one "rigid body", see below) controlled by the user, we did not use a physical environment to represent the virtual cube. The use of such a phantom would indeed limit the moves of the rigid body to translations and rotations from a pivot point (the entry point of the phantom). Intermediate points between the beginning of the path and the target are displayed along a certain distance from the current position of sight limited by a vision cutoff. The cutoff function is a steep sigmoid with inflexion point at 2 cm (the users may perceive about 2 to 4cm from the current position depending on their sensitivity). The intensity of points comprised between 0 (not visible) and 1 (intense) corresponds to levels of gray (between black and white) on the VDU and to voltages (comprised between 0 and about 10 volts) on the TDU [10, 12].

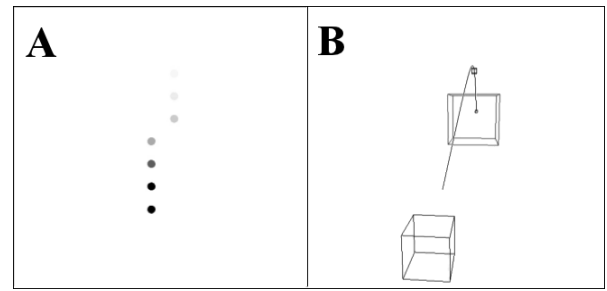

Figure 1. **(A)** Lateral view of the light pathway (path 1, see Fig. 2) as it is shown on the VDU in a 12x12 viewport. In the figure, the levels of gray are inverted (black corresponds here to the maximum intensity) **(B)** The same pathway viewed in high resolution. The beginning and the target correspond to the large squares. The small square represents the position at which the intensity of the points is equal to 0 and the trajectory no more perceptible.

Two different pathways are used: a curved one (Figs. 1 and 2) and a helical one (Fig. 3).

### C. Motion control – Trajectory records

Two input interfaces were used to control the position and the orientation of the virtual camera in order to follow the displayed pathway: (i) a keyboard to control the moves (forward and backward) and a mouse to control the orientation, or (ii) a "rigid body" tracked by the Polaris. This device, commonly used in computer assisted surgery [3], is an optical localizer that tracks in real time both the 3D position of 3 reflecting points on a *reference* rigid body and reflecting 3 points on a *moving* rigid body controlled by the user with his hand. By moving the current position and orientation of the virtual camera either with the keyboard and mouse or with the rigid body, the user draws a trajectory in the 3D space. The points of these trajectories are recorded every 5ms. Average correlations (in X and Y) along the Z axis (the calculus of such correlations are equivalent to time correlations but the time is replaced by the progression along the z axis) are computed to compare these series to the theoretical pathways. Standard deviations in X and Y are computed along the Z axis and their mean calculated. Correlations provide information about the similarity between the paths and the recorded trajectories while the average standard deviations provide information on the average distance between them, the average being calculated over all the deviations along the Z axis. The transit time, *i.e.* the time necessary to move from the beginning of the path to the target is also measured.

*D. Experiment*

To test the experimental platform, preliminary experiments were realized on 2 subjects, one naive (no prior use of a TDU) and one expert. In TDU mode, the subjects were blindfolded. They were instructed to track a trajectory in the most precise way and as fast as possible using one of the two input interfaces described above at once. The subject perceived the theoretical pathway from one of the modalities VDU or TDU. Each subject realized the same task 5 times in the four conditions 'VDU + keyboard + mouse', 'VDU + Polaris + Rigid body', 'TDU + keyboard + mouse', 'TDU+Polaris+Rigid body', and for each pathway (*i.e.* a total of 40 trajectories per user). A single session was programmed for each subject. The impact of training with multi-sessions will be studied later.

The experimental task was defined as follows. A training of the different controls and modalities of vision is practiced before starting the experiment. The subject has to move in the three dimensional space using the mouse and the keyboard, all directions being allowed. When the experiment begins, a timer is started and the subject controls its trajectory from the start to the target. The timer is stopped and the trajectory recorded once the target is reached. In the case of the keyboard and mouse control, the virtual camera that gives a view to the 3D environment is located at the starting point and oriented vertically in the bottom direction. In the case of the Polaris and rigid body control, the user holds the needle in his hand in vertical position, and starts from a reference point indicated by the experimenter.

III. RESULTS

Figures 2 and 3 show the average trajectories of the two users (each average trajectory is a mean of 10 individual records) in each condition. Table 1 gives quantitative comparisons between the trajectories and the theoretical paths, *i.e.* the average (along Z) standard deviations in X and Y from the theoretical path, and the correlations along the Z axis. One can observe that the average trajectories followed by the subjects are very close to the theoretical path, in all experimental conditions and for each path. This is confirmed with the very low standard deviations and very hight correlation levels obtained in most cases (table 1).

Some differences between the experimental conditions appear clearly in Table 1. Correlations are higher and standard deviations lower when the subjects perceive through the VDU than with the TDU. Nevertheless, correlation levels of about 55% are obtained with the TDU and the average standard deviations are comprised between 5mm (keyboard and mouse control) and 2cm (Polaris control).

Finally, it appears that the control of the trajectory is largely better when keyboard and mouse are used than when one use the Polaris and the rigid body held by the user's hand. Trajectories fluctuate more around the paths and the standard deviations are largely higher in this case.

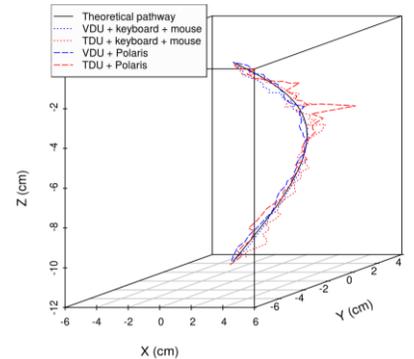

Figure 2. Path N°1 and the 4 average trajectories corresponding to the experimental conditions. The solid black line corresponds to the path.

We also observed that there were no significant differences between the precision of the trajectories of the naive subject compared to those of the expert. This is encouraging because it would mean that taking this system in hand may be very fast. Nevertheless, since it is difficult to conclude on the effect of expertise on the precision by comparing only two subjects, this result is not shown.

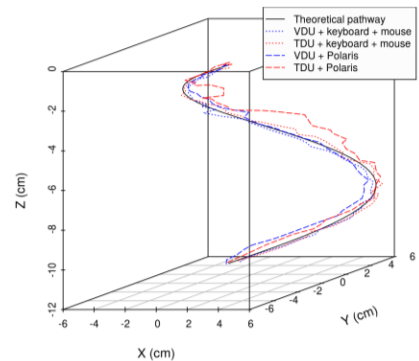

Figure 3. Path N°2 and the 4 average trajectories corresponding to the 4 experimental conditions. The solid black line corresponds to the path.

TABLE I. AVERAGE STANDARD DEVIATIONS OF TRAJECTORIES COMPARED TO THE THEORETICAL PATHWAYS, AND CORRELATIONS OF TRAJECTORIES WITH THE PATHWAYS ALONG THE Z AXIS. THE TRAJECTORIES OF THE NAIVE USER AND THE EXPERT BEING CLOSE TOGETHER (RESULT NOT SHOWN), THEY WERE USED AS ONLY ONE SET TO COMPUTE THE SD AND CORRELATION.

|  | Path | Average SD (cm) | Correlation (%) |
|---|---|---|---|
| VDU + Keyboard + Mouse | 1 | 0.03 | 87.3 |
|  | 2 | 0.08 | 83.0 |
| VDU + Polaris + Rigid Body | 1 | 0.18 | 83.6 |
|  | 2 | 0.33 | 90.4 |
| TDU + Keyboard + Mouse | 1 | 0.42 | 43.6 |
|  | 2 | 0.83 | 75.9 |
| TDU + Polaris + Rigid Body | 1 | 1.94 | 58.1 |
|  | 2 | 1.58 | 53.5 |

The average transit times are given in Table 2. The first observation is that the use of the TDU takes on average 4-5

times longer than when the paths are followed with the VDU. In addition, there seems to be an effect of the level of expertise on the speed of progression along the pathways. Finally, the complexity of path N°2 compared to path N°1 may also have an effect on transit times.

TABLE II. COMPARISON OF TRANSIT TIMES IN THE DIFFERENT CONDITIONS BETWEEN THE NAIVE SUBJECT AND THE EXPERT.

|  | Path | Naive subject | Expert |
|---|---|---|---|
| VDU + Keyboard + Mouse | 1<br>2 | 14.2 ± 9.3 s<br>18.2 ± 7.5 s | 8.6 ± 1.2 s<br>16.4 ± 2.6 s |
| VDU + Polaris + Rigid Body | 1<br>2 | 29.2 ± 3.7 s<br>33.0 ± 14.0 s | 16.4 ± 4.5 s<br>38.0 ± 6.3 s |
| TDU + Keyboard + Mouse | 1<br>2 | 52.0 ± 9.5 s<br>88.0 ± 16.1 s | 41.9 ± 22.4 s<br>51.2 ± 5.7 s |
| TDU + Polaris + Rigid Body | 1<br>2 | 73.1 ± 10.6 s<br>76.2 ± 19.6 s | 41.6 ± 25.2 s<br>55.0 ± 28 s |

IV. DISCUSSION

In this article, we presented a new manner to display the environment in a vision substitution device in order to navigate in a 3D environment without seeing with the eyes. Although preliminary, we obtained promising results. The observation that both an expert and a naive subject were able to follow very efficiently a light path in 3-dimensions by using a vision substitution device constitutes the main result of this work. Their trajectories were precise although the paths are small (12cm height and about 6cm of eccentricity) and twisted. Moreover, both subjects were able to take the system in hand easily although the level of expertise may have an influence on the speed of progression along the path whatever the vision mode or the control used.

Another result that should be highlighted is that the transit times with the TDU are 4-5 times longer than with the VDU. The reason is probably related to the cognitive load required to process visual information substitution [14]. The cognitive ressources necessary to execute the task may indeed increase in vision substitution as already shown in [12, 4] where response times were shown to increase when a TDU is used instead of a VDU.

Our set-up still suffers from a lack of an efficient control system. Although efficient, the keyboard and mouse control is not of natural use. Moves are precise because the TDU does not allow seeing far away from the current position and the user proceeds by small adjustments with the keyboard and the mouse. On the contrary, the Polaris and its rigid body held in the hand of the user are not comfortable to use. Without point of support or pivot (use of a physical phantom) the hand of the user is not stable and moves continuously by drift and small trembling. This induces an important decrease in the precision of positioning and orientation of the rigid body. This effect should be negligible in larger environments (longer paths, streets ...) or be considerably reduced by using a peripheral input more appropriate to small and precise moves (with many degrees of freedom) like a SPIDAR [15] or an endoscopic camera. We are now designing an indoor experiment where the position and posture of subjects are captured by stereoscopic cameras and the subjects perceive the environments composed of paths and objects, move in it and execute daily living tasks (take an object, put it elsewhere ...). Such an experiment should allow us to test the ability of subjects to navigate in a large 3D environment with a vision substitution system. This may also be adapted to prosthetic vision for blind people or used in augmented reality applications.

To conclude, let us remind that a question remains unanswered: we still do not know if the subjects are able to see in 3D with a vision substitution system and to understand this third dimension. The present experiment was not designed to test this assumption: to move along the path, the subjects can simply follow the light points one after the other in order of appearance. We however observed that when moving accidently far from the path, the subjects, young and healthy, were always able to come back to the path and continue. New experiments should nevertheless be designed to further clarify this point.